\documentclass[aps,prl,twocolumn,superscriptaddress,showpacs,floatfix]{revtex4-2}
\usepackage[english]{babel}
\usepackage{comment}
\usepackage{physics}
\usepackage{amsmath,amsthm,amssymb,amsfonts, color, comment, graphicx}
\usepackage{xcolor}
\usepackage{float}
\usepackage{graphicx}
\usepackage{dcolumn}
\usepackage{bm}
\usepackage{subfigure}

\UseRawInputEncoding
\usepackage{pgffor}
\usepackage{pdfpages}
\makeatletter
\AtBeginDocument{\let\LS@rot\@undefined}
\makeatother
\usepackage{hyperref}

\usepackage[normalem]{ulem}

\newcommand{\bea}{\begin{eqnarray}}
\newcommand{\eea}{\end{eqnarray}}

\newcommand{\br}{\mathbf{r}}

\newcommand{\be}{\begin{equation}}
\newcommand{\ee}{\end{equation}}

\newcommand{\bQ}{{{\bf{Q}}}}

\newcommand{\beal}{\begin{align}}
\newcommand{\eeal}{\end{align}}
\newcommand{\ra}{\rangle}
\newcommand{\la}{\langle}

\newcommand{\upa}{\uparrow}
\newcommand{\dna}{\downarrow}

\newcommand{\pdg}{{\phantom\dagger}}

\newcommand{\btjstrw}{\mathrel{{\rotatebox[origin=c]{90}
{$\bowtie$}}\kern-0.18em\raisebox{-.95ex}{$\bullet$}
\kern-0.5em\raisebox{.97ex}{$\bullet$}
\kern-1.12em\raisebox{.97ex}{$\bullet$}
\kern-0.52em\raisebox{-.95ex}{$\bullet$}}}

\newcommand{\btjnbrR}{{\mathrel{\rotatebox[origin=c]{90}
{$\bowtie$}}\kern-0.22em\raisebox{.9ex}{$\bullet$}
\kern-1.em\raisebox{-.8ex}{$\bullet$}}}
\newcommand{\btjnbrL}{{\mathrel{\rotatebox[origin=c]{90}
{$\bowtie$}}\kern-0.22em\raisebox{-.8ex}{$\bullet$}
\kern-1.em\raisebox{+.9ex}{$\bullet$}}}

\def\a{\alpha}
\def\b{\beta}

\def\e{\epsilon}

\def\g{\gamma}

\def\m{\mu}

\def\s{\sigma}

\def\x{\xi}

\newcommand{\llangle}[1][]{\savebox{\@brx}{\(\m@th{#1\langle}\)}%
  \mathopen{\copy\@brx\kern-0.5\wd\@brx\usebox{\@brx}}}
\newcommand{\rrangle}[1][]{\savebox{\@brx}{\(\m@th{#1\rangle}\)}%
  \mathclose{\copy\@brx\kern-0.5\wd\@brx\usebox{\@brx}}}

\begin{document}

\preprint{APS/123-QED}

\title{Searching for superconductivity in doped triangular lattice Kitaev magnets}
\author{Andrew Hardy}
\thanks{andrew.hardy@mail.utoronto.ca}
\affiliation{Department of Physics, University of Toronto, 60 St. George Street, Toronto, ON, M5S 1A7 Canada}
\author{Ryan Levy}
\thanks{Current affiliation: PsiQuantum, 700 Hansen Way, Palo Alto, CA 94304, USA } 
\affiliation{Center for Computational Quantum Physics, Flatiron Institute, New York, New York 10010, USA}
\author{Arun Paramekanti}
\affiliation{Department of Physics, University of Toronto, 60 St. George Street, Toronto, ON, M5S 1A7 Canada}
\date{\today}

\begin{abstract}
Motivated by exploring correlated metals with frustrating bond-dependent exchange interactions, we study hole and electron
doped Kitaev Mott insulators on the triangular lattice.
Using homogeneous parton mean field theory, we find that the stripe antiferromagnetic (AFM) order for Kitaev coupling $K>0$ and the ferromagnetic (FM) order for $K<0$, both vanish at sufficiently large doping, beyond which we find regimes with chiral $d\pm i d$ singlet pairing and $p\pm ip$ triplet pairing respectively.
Our tensor network computations however reveal that the superconducting correlations are strongly suppressed; while FM order stubbornly persists for the doped $K<0$ model, the doped $K>0$ model features emergent spin-charge modulated stripe orders.
At higher hole doping for $K > 0$, where AFM order is more strongly suppressed than for the electron doped case,
incorporating a sufficiently strong nearest-neighbor attraction yields evidence for singlet $d$-wave superconductivity  with Luttinger parameter $K_{\rm sc} < 1$.
Our work sets the stage for a broader exploration of doping effects in triangular lattice magnets such as NaRuO$_2$ which
feature bond-dependent exchange interactions. 
\end{abstract}

\maketitle
\textit{Introduction.|}
Doping correlated magnetic insulators has long been considered a viable path to achieve high-temperature superconductivity (SC)  since  the discovery of the cuprate superconductors \cite{Anderson87,Emery87, Lee2006, Scalapino2012}. 
Based on the picture of resonating valence bond singlets in a Mott insulator as blocked Cooper pairs, parent quantum spin liquids \cite{Balents2010, Zhou2017} have been a favorite hunting ground for emergent SC upon doping which enables Cooper pair mobility.
Stabilizing a parent quantum spin liquid requires spin exchange frustration in the Mott insulator which may be induced by lattice geometry \cite{SavaryBalents2017}, multispin interactions near the Mott transition 
\cite{Cookmeyer201, Kuhlenkamp2024}, or bond-dependent interactions from spin-orbit coupling (SOC) as exemplified by the honeycomb Kitaev spin model 
\cite{Kitaev2006,Takagi2019,Trebst2022}.
This has opened up the broad search for SC in doped Mott insulators where lattice geometry and unconventional spin interactions  conspire to induce pairing. \par

For the honeycomb lattice Kitaev model, relevant to a wide class of spin-orbit coupled Mott 
insulators such as {Na$_2$IrO$_3$}, {Li$_2$IrO$_3$}, and {$\alpha$-RuCl$_3$} \cite{Takagi2019,Trebst2022,Matsuda2025}, 
the nature of the ground state induced by doping has proven difficult to determine.
While parton mean-field theory (MFT) approaches suggested that the doped Kitaev spin liquid could be an unconventional superconductor \cite{Burnell2011, You2012, Halasz2014, Liu2016}, tensor network studies show that the doped ferromagnetic (FM) Kitaev model displays enhanced Nagaoka-like ferromagnetism with no signs of SC \cite{Jin2024}.
The doped antiferromagnetic (AFM) Kitaev  model shows an exponential decay of all superconducting correlations, although pair-density waves appear as the leading pairing channel correlations \cite{ Peng2021a, Laurell2024}.
Frustrated triangular lattice magnets have also long been explored for unconventional SC upon doping \cite{Watanabe2004,Kyung2006,Zhu2022}. 
This has been inspired in part by early results of superconductivity in Na$_{1-x}$CoO$_2$ \cite{ Takada2003, Khaliullin2004}. 
Remarkably, experiments on Sn/Si(111) have found evidence for chiral $d \pm id$ SC \cite{Ming2023} on the triangular lattice, 
which appears to be captured by nonlocal extensions of dynamical mean field theory \cite{Cao2018,Ming2023}. 
Tensor network studies of superconductivity in the triangular lattice also find that pairing only occurs in the presence of additional frustration or longer-range interactions \cite{Jiang2021, Jiang2021b, Huang2022, Huang2023, Kiely2024}. 
These examples highlight the importance of advanced computational algorithms to complement mean field studies of strongly correlated electronic Hamiltonians.

\par
 
In this Letter, we explore the electronic ground state resulting from doped mobile charge carriers in  the  triangular lattice Kitaev model.
The triangular lattice Kitaev spin model has a magnetically ordered ground state, exhibiting FM order for Kitaev exchange $K < 0$ and
stripe AFM order for $K>0$ \cite{Maksimov2019}.
On the materials front, a stripe-ordered ground state driven by bond-anisotropic exchanges has been recently discovered in ${\mathrm{CsCeSe}}_{2}$ \cite{Xie2024}.
Such bond-anisotropic terms also appear in descriptions of rare earth magnets such as YbMgGaO$_4$ 
and related materials \cite{Li2016, Zhu2017, Sanders2017, Luo2017, Cevallos2018, Baenitz2018, Kimchi2018, Zhu2018, Maksimov2019},
as well as in the more recently studied ruthenate
NaRuO$_2$ \cite{Ortiz2023, Razpopov2023, Bhattacharyya2023} which features $j_{\text{eff}} = 1/2$ moments from
the Ru $d^5$ configuration coupled via an AFM Kitaev exchange term \cite{Ortiz2023, Razpopov2023, Bhattacharyya2023} 
in addition to other exchange interactions.
However, the broad phenomenology of doped Kitaev magnets on the triangular lattice has not been explored to the best of our knowledge,
making it an important model to explore the impact of doping. \par  

Here, we study the doped triangular lattice Kitaev model using parton MFT \cite{Coleman2015} and tensor network computations using variationally uniform matrix product states (VUMPS) \cite{Zauner-Stauber2018, Vanderstraeten2019} which provides the thermodynamic ground-state of the quantum system in a quasi one-dimensional regime. 
Our parton MFT shows that magnetic order of the Mott insulator persists up to a critical doping, beyond which the ground state is an unconventional  $d$-wave singlet ($K > 0$) or $p$-wave triplet ($K < 0$) superconductor. 
By contrast, while our VUMPS simulations confirm the magnetic orders found in MFT at half-filling, they instead show that the doped state exhibits persistent FM order ($K<0$) or emergent spin-charge stripe orders ($K>0$) which strongly suppresses pairing. 
These spin-charge coupled stripe orders found in our work appear as holes doped into antiphase domain walls in the parent stripe AFM order.

Upon further doping, we find that the spin stripe order is lost, giving way to a non-superconducting state with a spin gap but 
no charge gap. 
This raises the intriguing possibility that this regime in the $2$D limit could potentially be a fractionalized algebraic charge liquid with gapped bosonic spinons and mobile fermionic holons \cite{Kaul2008}.
\par
Since our VUMPS simulations show that superconductivity seems to be absent in the pure doped Kitaev model, we analyze the impact of an additional nearest-neighbor attractive interaction, possibly induced by phonons.
For a sufficiently strong attractive interaction, we find evidence for induced $d$-wave pairing. 
Our results may have broader implications for phonon-induced long-range attraction \cite{Wang2021b, Qu2022, Wang2025c} which could be relevant to extended Hubbard models for moir\'e transition metal dichalcogenides \cite{Zhou2022, Chen2023b}, and experiments on such materials \cite{Ciorciaro2023}.
Our work highlights the importance of newly developed numerical VUMPS algorithms, demonstrating its strengths and value in complementing MFT 
with careful studies of SC and competing orders in strongly correlated electronic systems with gapless phases. \par

\textit{Model Hamiltonian.} The physics of doped Mott insulators relies on the competition between magnetic exchanges and kinetic energy. 
We begin with the Kitaev spin model on the triangular lattice in the undoped limit
\begin{equation}
\begin{gathered}
\label{eq:Kitaev}
    H_{K}= K    \sum_{\langle \br, \br' \rangle} S^{\mu_{\br \br'}} (\br) S^{\alpha_{\br \br'}} (\br')
\end{gathered}
\end{equation}
with $\mu_{\br\br'} = \{x,y,z \} $ for the three bond directions.
These directions are $\hat{x} \!=\! (\frac{1}{2}, -\frac{\sqrt{3}}{2}), \hat{y} \!=\! (\frac{1}{2}, \frac{\sqrt{3}}{2}), \textrm{and } 
\hat{z} \!=\!  (1,0)$.
The classical ground state for $K >  0$ is a nematic with broken spin-lattice rotation symmetry, featuring pockets of stripe AFM order along 
various directions without long-range magnetic order \cite{Catuneanu2015}. 
The $K <  0$ classical ground state lies at the boundary of FM and stripe AFM orders \cite{Catuneanu2015}. 
In the quantum limit, previous studies considered the possibility of a chiral spin liquid phase for the $K  >  0$ case  \cite{Li2015}, although recent consensus is that the ground state is magnetically ordered: 
a period-2 stripe AFM for $K  >  0$ and a ferromagnet for $K  <  0$ \cite{Becker2015, Li2016, Maksimov2019, Wang2021a}. 
\par

To include mobile charge carriers in the system, we introduce an isotropic kinetic hopping term parameterized by $t$ terms, as a generalization of the  $t$-$J$ model. 
\begin{equation}
\begin{gathered}
    H=-t\sum_{{\langle \br, \br' \rangle}} \mathcal{P}\left(c^{\dagger}_{\br, \a} c^\pdg_{\br', \a}+\mathrm{h.c.}\right) \mathcal{P}+ H_{K},
    \label{eq:gutzwillerhamiltonian}
\end{gathered}
\end{equation}
where $\mathcal{P}$ is the Gutwziller projection that projects out doubly occupied sites to account for
the large local Hubbard repulsion. 
We fix $t=1$ and tune the ratio $K/t$ and dopant concentration $\delta = 1 - n$, with $\delta > 0$ and $\delta <0$ corresponding to hole and electron doping respectively. \par

\begin{figure}[t]
     \centering
        \includegraphics[width=0.49\textwidth]{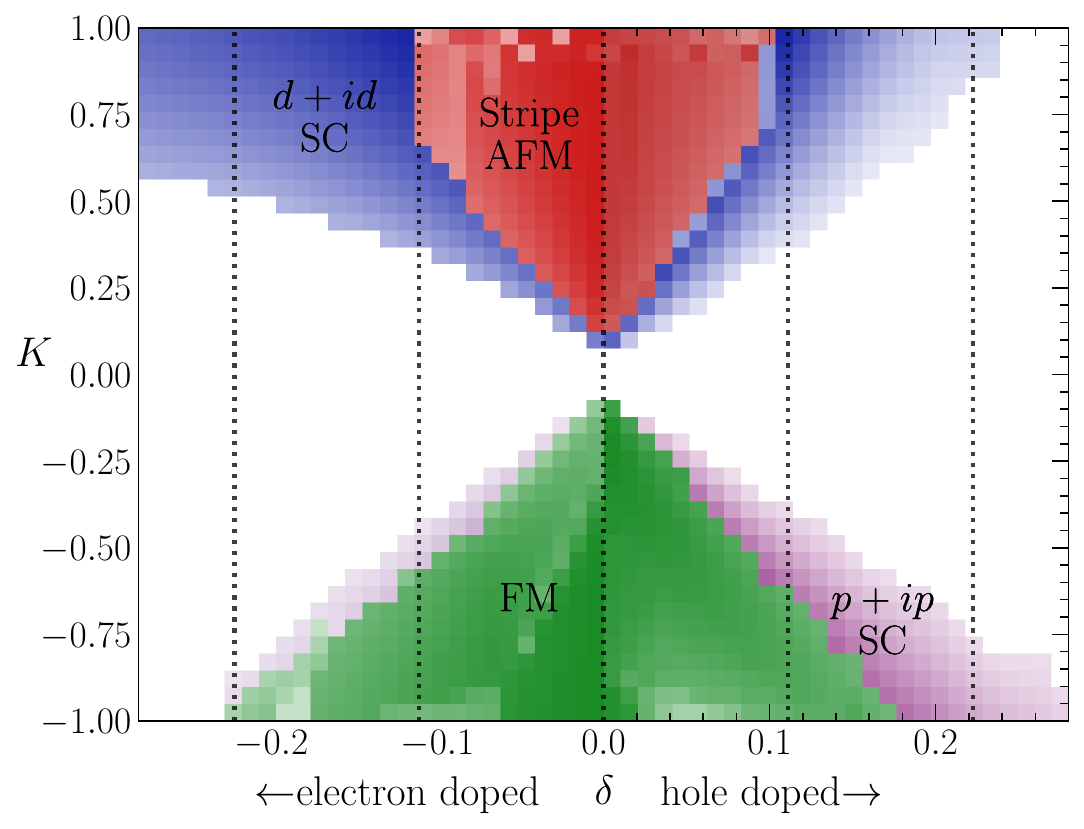}
    \caption{Phase diagram of doped Kitaev model ($t$-$K$ model) on the triangular lattice from homogeneous parton mean field theory 
    as a function of doping and Kitaev exchange $K$ (in units where
    $t =1$). Here $\delta > 0$ ($\delta <0$) represents hole (electron) doping. For AFM exchange $K > 0$, we find time-reversal 
    broken $d+id$ SC at the
    boundary of period-two stripe AFM order. 
    For FM exchanges ($K < 0$), $p+ip$ wave SC emerges at the phase boundary where FM order vanishes. 
    Color saturation represents the relevant order parameter.
    Vertical cuts represent the scans performed with tensor network methods. }
    \label{fig:kitaev_sbmft_sc}
\end{figure}
\textit{Parton Mean Field Theory.---} 
To gain a qualitative picture of the stability of the magnetic states or the possibility of novel charge states, 
we consider a homogeneous parton mean-field theory of the $t$-$K$ model.
We consider a parton decomposition, otherwise known as slave-boson theory \cite{Coleman2015}, $ c^\dagger_{\br,\s} = b^\pdg_\br f^\dagger_{\br,\s}$, with the spin operator then represented as
$S^\mu(\br) = \frac{1}{2} f^\dagger_{\br\alpha} \s_{\alpha\beta}^\mu f^\pdg_{\br'\beta}$.
The details of the implementation are presented in the appendix. \par
The mean field results demonstrate several striking phases shown in Fig.~\ref{fig:kitaev_sbmft_sc}. At zero doping, we recover the FM or stripe AFM
order. 
Upon doping with holes or electrons, the magnetic orders remain robust, and persist up to a critical doping
$\delta_c = \gamma |K|/t$ with $\gamma  \approx  0.15$-$0.2$.
At the boundaries of these magnetic phases, for both hole and electron doping, we find emergent topological SC, with $(d+id)$ singlet for $K > 0$ and $(p +ip) |\upa\upa  +  \dna\dna \rangle$ triplet for $K < 0$.  
These superconducting results are consistent with earlier 
work \cite{Li2015}, although we include the possibility of competing magnetic orders, which was not previously considered.
Details of the pairing state are given in the appendix. 
Over the indicated doping range, the phase diagram exhibits only weak particle-hole asymmetry which we attribute to the asymmetry of the density of states on the triangular lattice.
These results provide a fascinating example of emergent topological SC at the boundary of frustrated magnetic states,
and warrant further investigation using numerical techniques capable of handling the strong correlation constraint.

\par

\textit{Variational Uniform Matrix Product States.|}
\begin{figure*}[ht]
     \centering
        \includegraphics[width=0.49\textwidth]{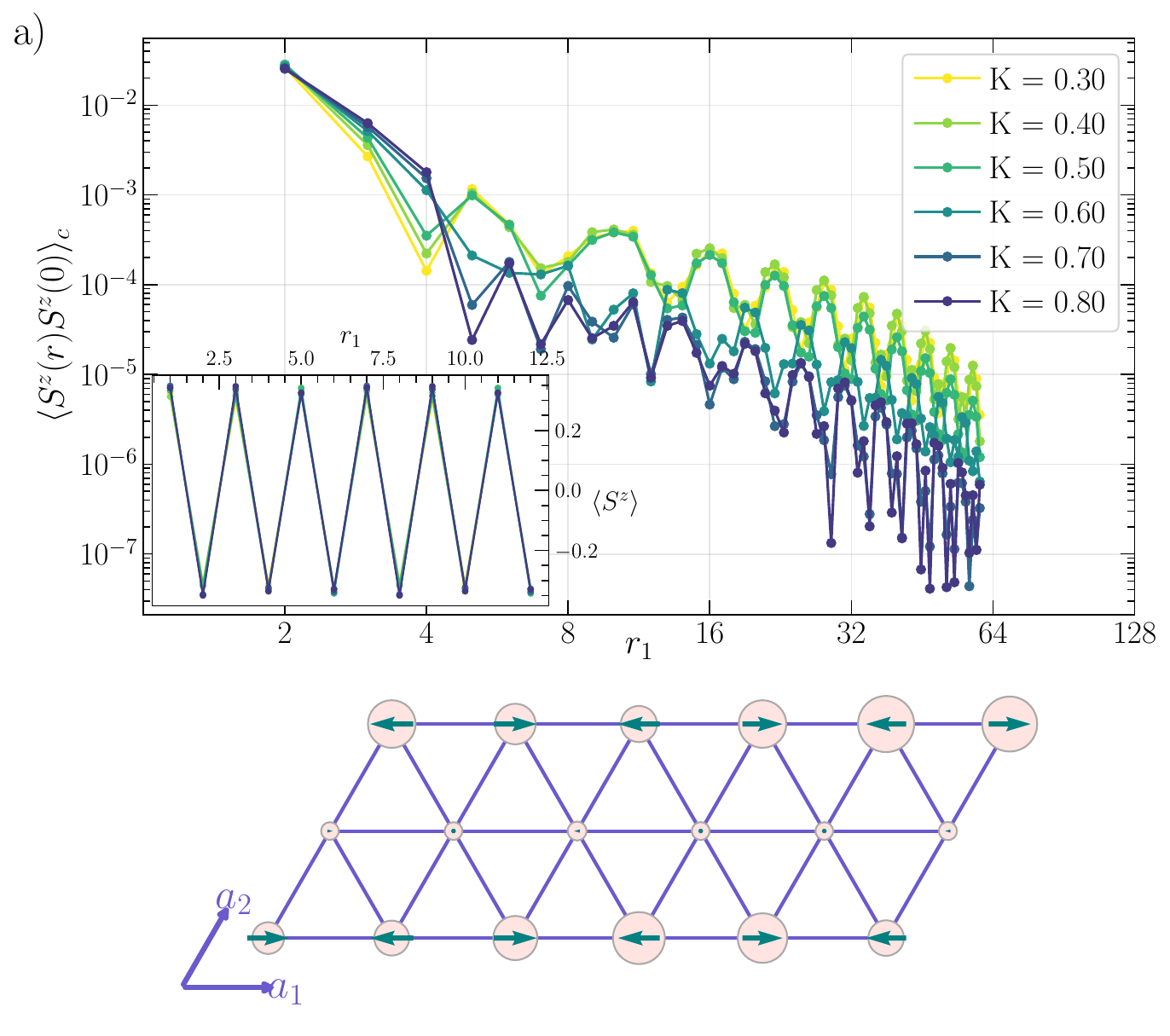}
        \includegraphics[width=0.49\textwidth]{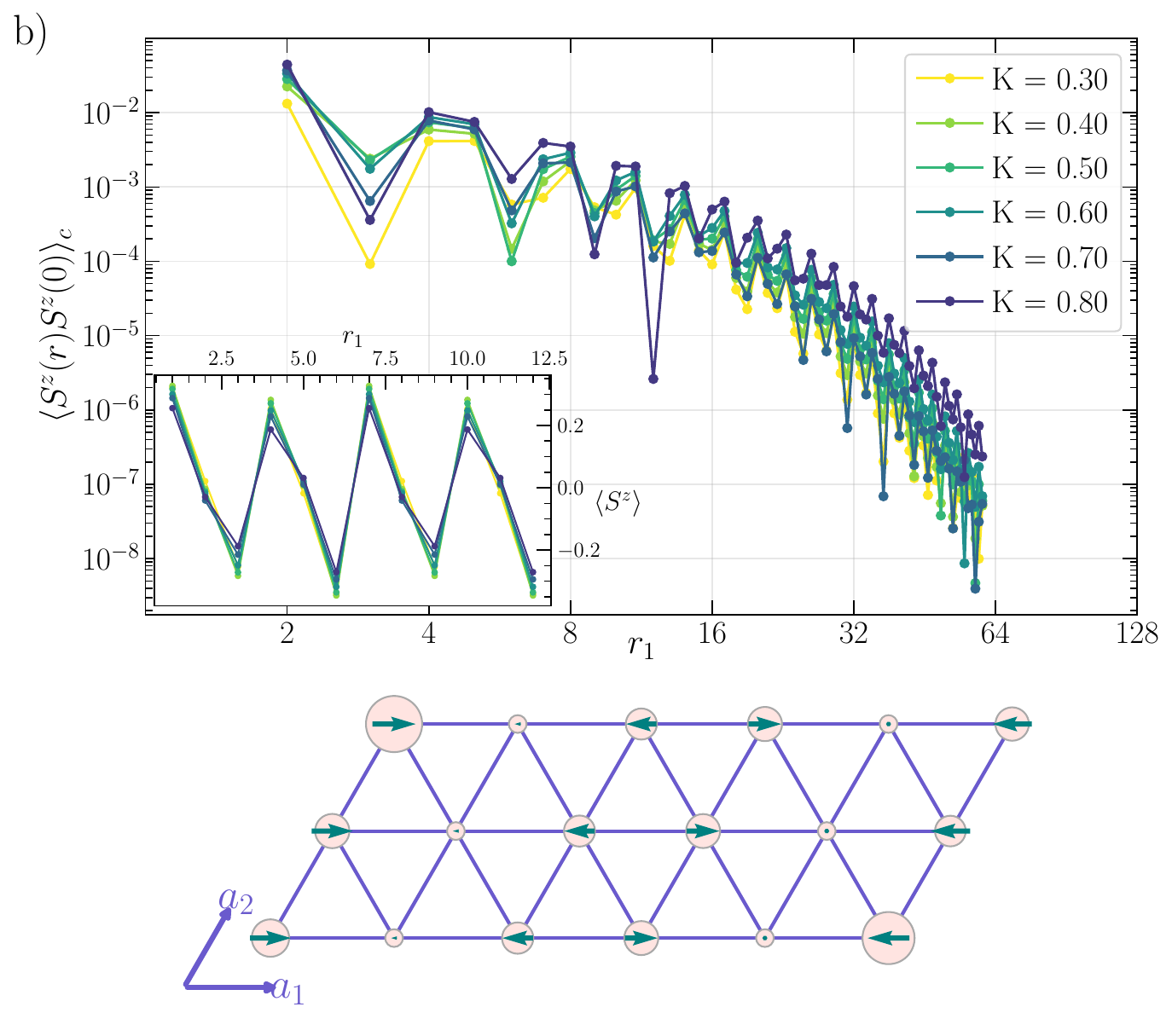}
    \caption{Connected spin correlations as a function of distance along the leg $r_1$ for the doped $t$-$K$ model with varying $K>0$ (in units where $t=1$)
    for (a) hole doping $\delta=1/9$ and (b) electron doping $\delta=-1/9$. Lower panels show the $6\times 3$ unit cell with 
    $\la S^z(\br)\ra$ indicated by arrows and charge density $\la n(\br)\ra$
    depicted by the size of the circles. Insets in the top panels show the spin expectation
    value $\la S^z(\br)\ra$ on the lowest leg of the unit cell as a function of $r_1$. These results are consistent with spin-charge modulated
    stripe order for both cases, but with distinct ordering wavevectors, and enhanced spin fluctuations at smaller $K$ for hole doping.}

    \label{fig:oneninth}
\end{figure*}
To determine the validity of the parton MFT SC state,
we explore the undoped and weakly doped ground state using
variational uniform matrix product states.
Similar to the infinite-density matrix renormalization group (iDMRG), the VUMPS algorithm performs a variational search over a space of quantum wave functions represented by a uniform matrix product states (UMPS) which capture the thermodynamic limit in the longitudinal direction.
We find that finite DMRG tests on the triangular lattice suffer from severe finite size and boundary effects, which motivates studying the system in the infinite limit.
Calculations in the thermodynamic limit also have several practical and conceptual advantages in computing quantities such as the transfer matrix and correlation functions \cite{Haegman2016, Zauner-Stauber2018, Vanderstraeten2019}. 
The VUMPS algorithm optimizes directly rather than using a self-consistent `bath' routine like iDMRG, which has been shown to speed up convergence 
in gapless systems \cite{Zauner-Stauber2018,Kiely2022}. 
 Details about the algorithmic implementation are in the appendix.
\par 

Using VUMPS, we have explored several cuts in the phase diagram from Fig.\ \ref{fig:kitaev_sbmft_sc} for hole and electron doping and both signs of $K$.
Commensurability within the unit cell limits our explored doping values to $\delta   =   \pm 1/9,\pm 2/9$ for computationally accessible $18$-site unit cells.
We have studied both $9 \times   2$ and $6 \times  3$ unit cells, but we focus here on results in the latter geometry (see Fig.\ref{fig:oneninth}) 
as it is more representative of the 2D limit.
As these system geometries remain in the quasi-$1$D limit where MPS methods remain effective, we report the direct ground state without extrapolating to the full $2$D thermodynamic limit.
To characterize the spin and charge orders, we compute  $\la S^\mu (\br) \ra$ and $\la n(\br) \ra$ as well as the connected correlations 
$\langle S^\mu (\br) S^\mu{({\bf 0})} \rangle^\pdg_c  \equiv  \langle S^\mu(\br) S^\mu ({\bf 0}) \rangle  -  \langle S^\mu (\br) \rangle \langle S^\mu (\bf 0) \rangle$ and
$\langle n(\br) n({\bf 0}) \rangle^\pdg_c \equiv \langle n(\br) n({\bf 0}) \rangle - \langle n(\br) \rangle \langle n({\bf 0} )\rangle$. 
To compute the spin and charge order, we provide a weak symmetry breaking Weiss field along $S^z$ at a single site which is reduced to zero in the initial 
sweeps (with bond dimension $\chi < 64$) of the variational search which provide an ordering preference. 
We confirm that the undoped model has stripe-AFM order for $K  >  0$ and FM order for $K  <  0$. 

For nonzero dopings we have explored, we find that the ground state for $K <  0$ is a uniform FM metal with strongly suppressed pair correlations. 
We present pertinent results for that case in the Supplemental Material (SM) \cite{SuppMat}.
We focus below on results for $K >  0$, where we find emergent spin-charge orders and non-magnetic metallic states which might support SC.

{\it Weak doping of Kitaev AFM, $\delta = \pm 1/9$.|}
As shown in the lower panels and insets of Figs.~\ref{fig:oneninth} (a) and (b), a spin-charge stripe order emerges from weak doping of the Kitaev 
Mott insulator.
In the hole-doped case, Fig~\ref{fig:oneninth}(a), the spin order has wavevector $\bQ_{\rm sp}  \equiv  (Q_1,Q_2)  =  (\pi,2\pi/3)$, 
leading to charge modulation at $\bQ_{\rm ch}  =  2\bQ_{\rm sp}  =  (0,4\pi/3)$. 
This strong charge modulation has a $\sim  10\%$ reduction in the charge density in the central leg of the three-leg cylinder.
In addition, we find a weaker $\lesssim   1 \%$ charge modulation along the legs, with a 
wavelength equal to the unit cell, indicative of universal quasi-1D Luttinger charge-density-wave (CDW) behavior \cite{White2002}.
The electron-doped case, Fig~\ref{fig:oneninth}(b), also exhibits spin-charge stripes, with a distinct wavevector $\bQ_{\rm sp}  =  (2\pi/3,0)$ and 
charge modulation at $\bQ_{\rm ch} = 2\bQ_{\rm sp} = (4\pi/3,0)$. 
This leads to every third rung having a strongly depleted charge density as shown, about $\sim 10\%$ lower than adjacent rungs. 
Again, in this case, we observe a weaker superposed quasi-1D charge modulation along the legs.
In a manner similar to the spin-charge stripes seen in various hole-doped cuprates \cite{Kivelson1998,Fradkin2015,Tranquada2020_stripes}, 
we can view the strong spin-charge stripe 
orders in the weakly doped Kitaev Mott insulator as `rivers of charge' separating domains of stripe-AFM order.
The strength of the spin-stripe states with $\bQ_{\rm ch}  =  2\bQ_{\rm sp}$ for both hole and electron doping,
and their persistence irrespective of the lattice geometries explored, indicate that these are genuine ground state orders potentially
likely to persist in 2D,
while the quasi-1D modulations should weaken and disappear in the 2D limit \cite{Gannot2023}.

\begin{figure*}
     \centering
        \includegraphics[width=0.49\textwidth]{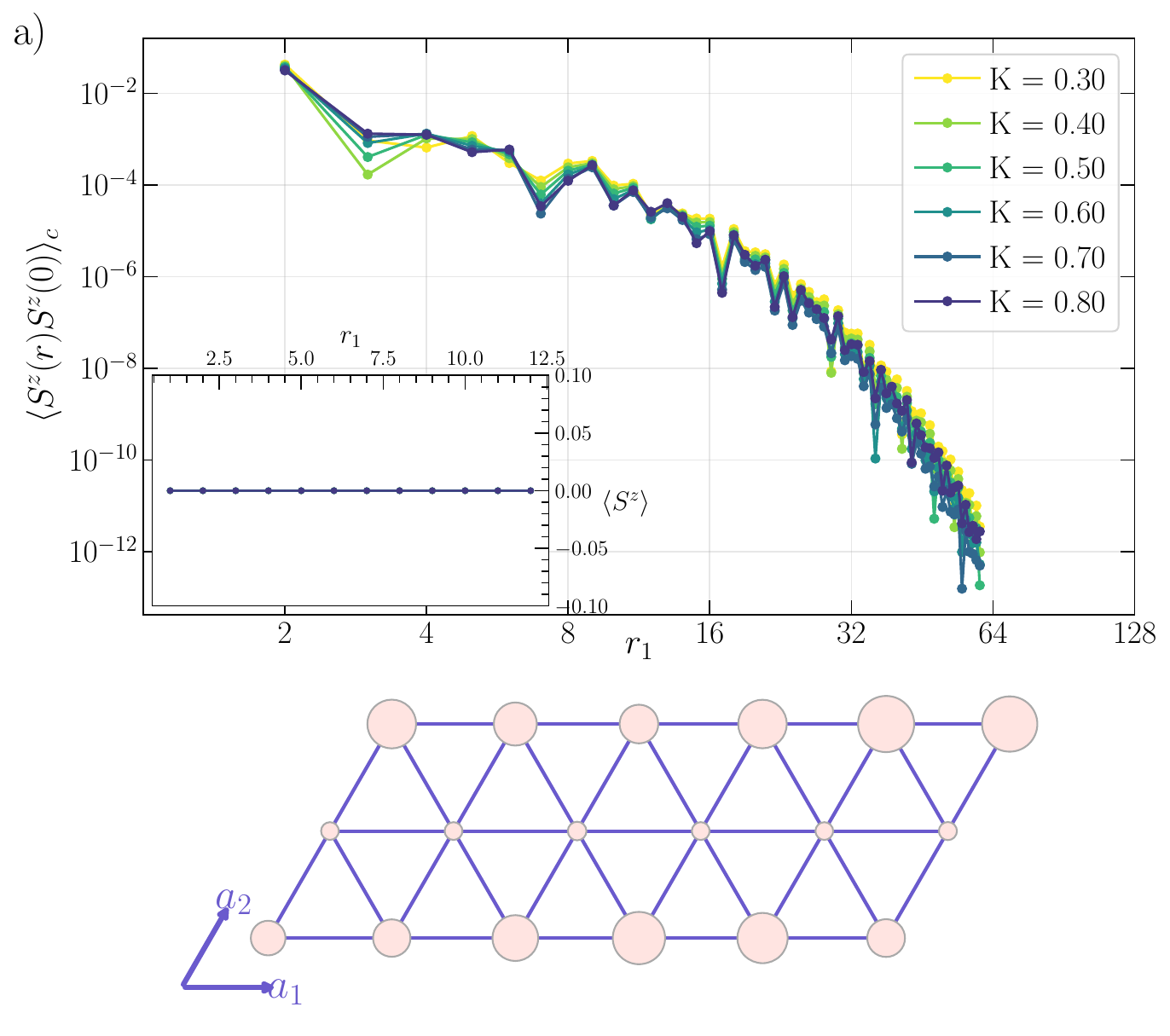}
        \includegraphics[width=0.49\textwidth]{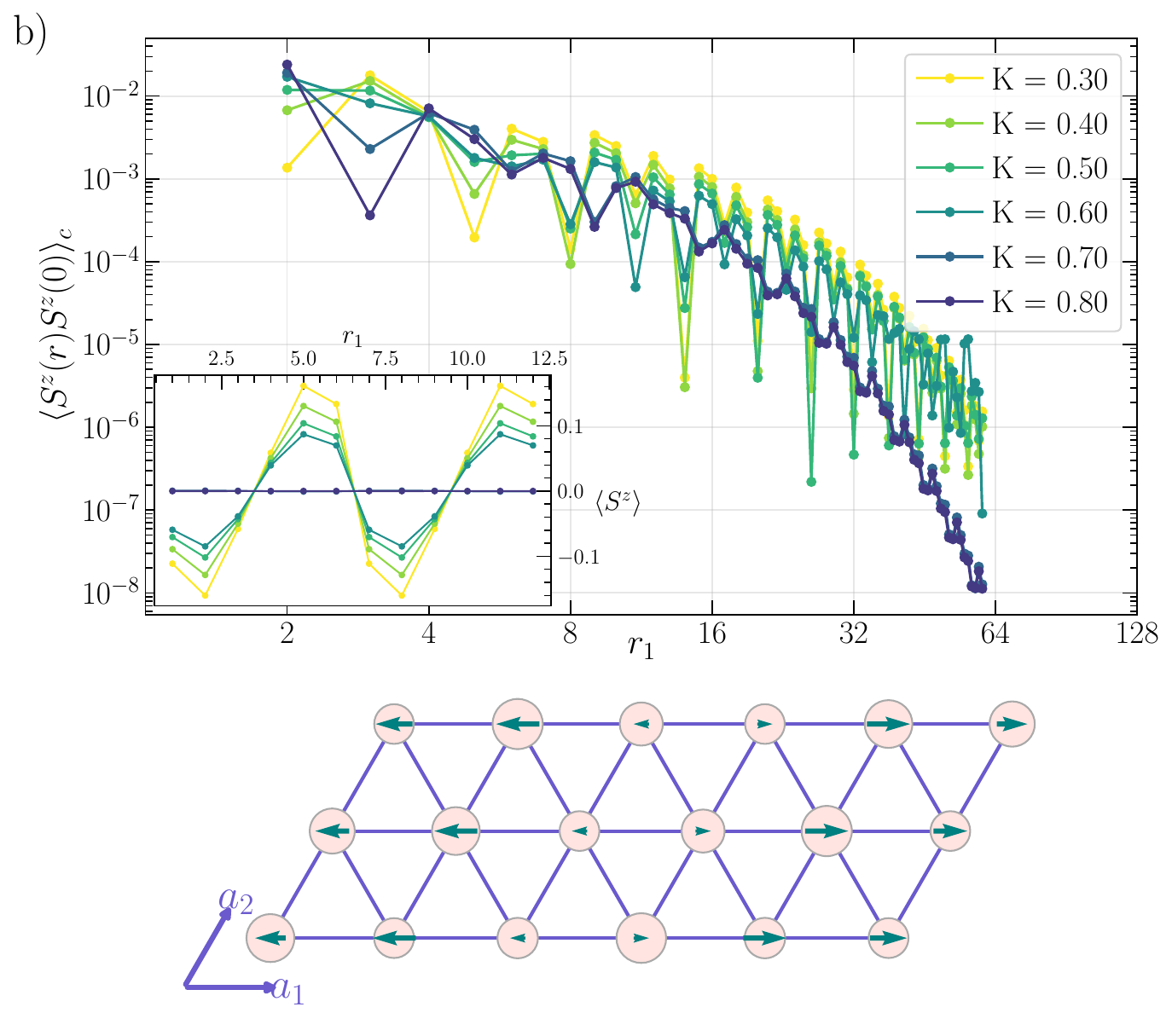}
     \caption{Connected spin correlations as a function of distance along the leg $r_1$ for the doped $t$-$K$ model with varying $K>0$ (with $t=1$)
    for (a) hole doping $\delta=2/9$ and (b) electron doping $\delta=-2/9$. Lower panels show the $6\times 3$ unit cell with 
    $\la S^z(\br)\ra$ indicated by arrows and charge density $\la  n(\br) \ra $
    depicted by the size of the circles. Insets in top panels show the spin expectation
    value $\la S^z(\br)\ra$ on the lowest leg of the unit cell as a function of $r_1$. These results are consistent with charge stripe order with
    no magnetism for $\delta=2/9$ and persistent spin-charge modulated stripe order with enhanced spin fluctuations for $\delta=-2/9$.}
    \label{fig:twoninth}
\end{figure*}

We next turn to the connected spin correlation function $\langle S^\mu(\br) S^\mu({\bf 0}) \rangle^\pdg_c$ shown in the main panels of 
Figs.~\ref{fig:oneninth} (a) and (b), which inform us about spin fluctuations on top of the ordered state.
The electron-doped case exhibits spin fluctuations which do not change much with $K$, while the hole-doped case exhibits enhanced spin
fluctuations as we decrease $K$, hinting at a magnetic critical point at this fixed doping.
Upon decreasing the Kitaev coupling from $K=0.9$ to $K=0.3$, the correlation length doubles to nearly $4$ unit cells ($24$ sites), although we do not access a gapless critical  point within our restricted sweep. 
The gapless regime $\abs{K} < 0.3 $, however, becomes difficult to converge within VUMPS. 
The nature of the transition from a gapless Fermi liquid to these spin and charge ordered states warrants further investigation, despite the difficulty of such critical states admitting a MPS representation. 
Such a magnetic critical point could also be accessed by increasing doping at fixed $K$. 

{\it Strong doping of Kitaev AFM, $\delta =  \pm 2/9$.|}
Upon further doping, as shown in Fig.~\ref{fig:twoninth}(a), the magnetic order is lost on the hole doped side with $\delta =  2/9$, 
while a charge modulation persists with the central leg having an $\sim 10\%$ depleted charge. 
This is consistent with charge stripe order without an accompanying spin stripe order.
In contrast to an SU(2) invariant spinful Luttinger liquid which exhibits long range power-law spin correlations, 
the spin correlations in this hole-doped Kitaev metal are strongly suppressed and appear short-ranged \ref{fig:twoninth}. 
On the electron-doped side, for $\delta =  -2/9$, as seen from Fig.~\ref{fig:twoninth}(b), the coupled spin-charge modulated stripe order persists, with evidence of enhanced spin fluctuations, so that even higher electron doping is needed to access a magnetic critical point.
Intriguingly, at this electron doping value, the residual spin order in the spin-charge stripe is seen to get stronger at smaller $K$,
highlighting the potential role of Haerter-Shastry type kinetic magnetism \cite{Haerter2005} in determining the spin ordering and particle-hole asymmetry seen in our numerical results. 
Previous work \cite{Sherif2025} in fact also sees stripe order in the strict $U = \infty$ limit that emerges upon sufficient hole doping of the $120^\circ$ order, due to kinetic magnetism effects.
\par

{\it Pair correlations in doped AFM, $\delta= 2/9$.|}
We now explore the possibility of unconventional superconductivity in these doped Kitaev-Mott insulators.
While the VUMPS calculations are performed in a U$(1)$ charge conservation sector, we can determine the superconducting susceptibility through pair correlation functions. 
Motivated by our parton MFT results, we compute the singlet pair-correlation function
$
P_s(\br)=\left\langle \Delta^{\dagger}_s(\br)\Delta_s(0)\right\rangle
$
with the singlet operator 
$
\Delta_s^{\dagger}(\br)=
(c_{\br, \downarrow}^{\dagger} c_{\br+1,\uparrow}^{\dagger}-c_{\br, \uparrow}^{\dagger} c_{\br+1,\downarrow}^{\dagger})/\sqrt{2}$.
We have similarly defined and computed triplet pair correlations, $P_t(\br)$, but find that these are much more strongly suppressed for both signs of $K$ 
and all dopings; we thus relegate these results to the SM \cite{SuppMat}.
\begin{figure}[htb]
     \centering
\includegraphics[width=0.49\textwidth]{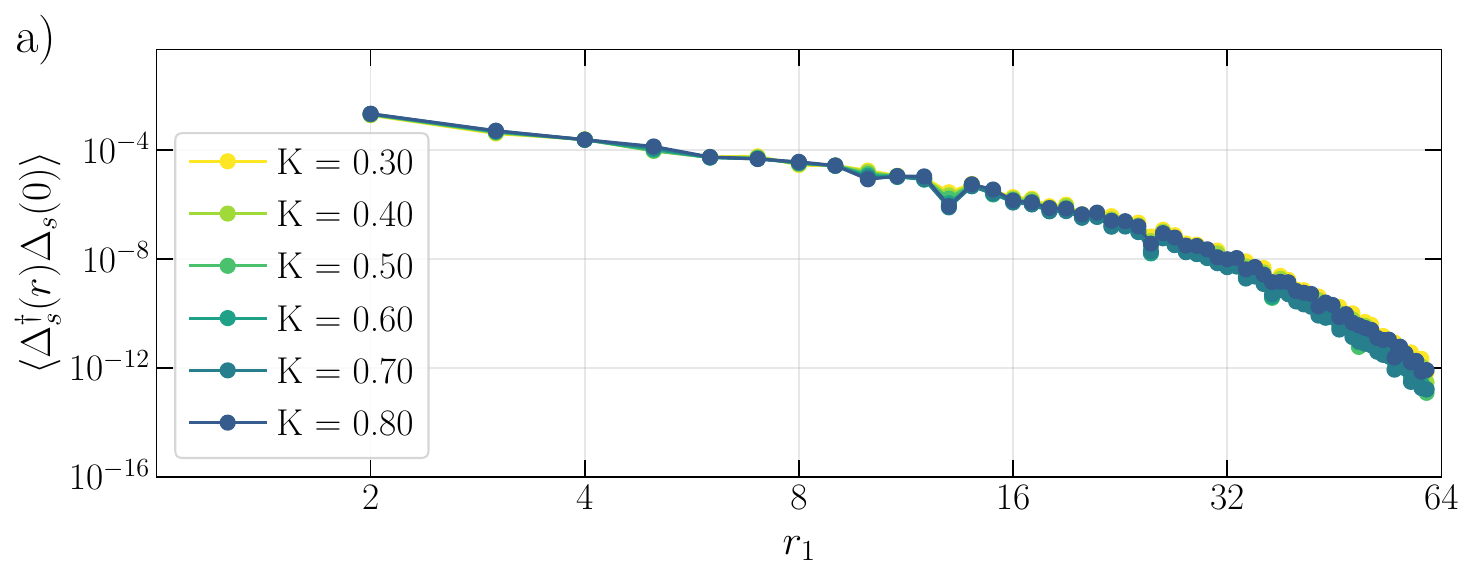}
\includegraphics[width=0.49\textwidth]{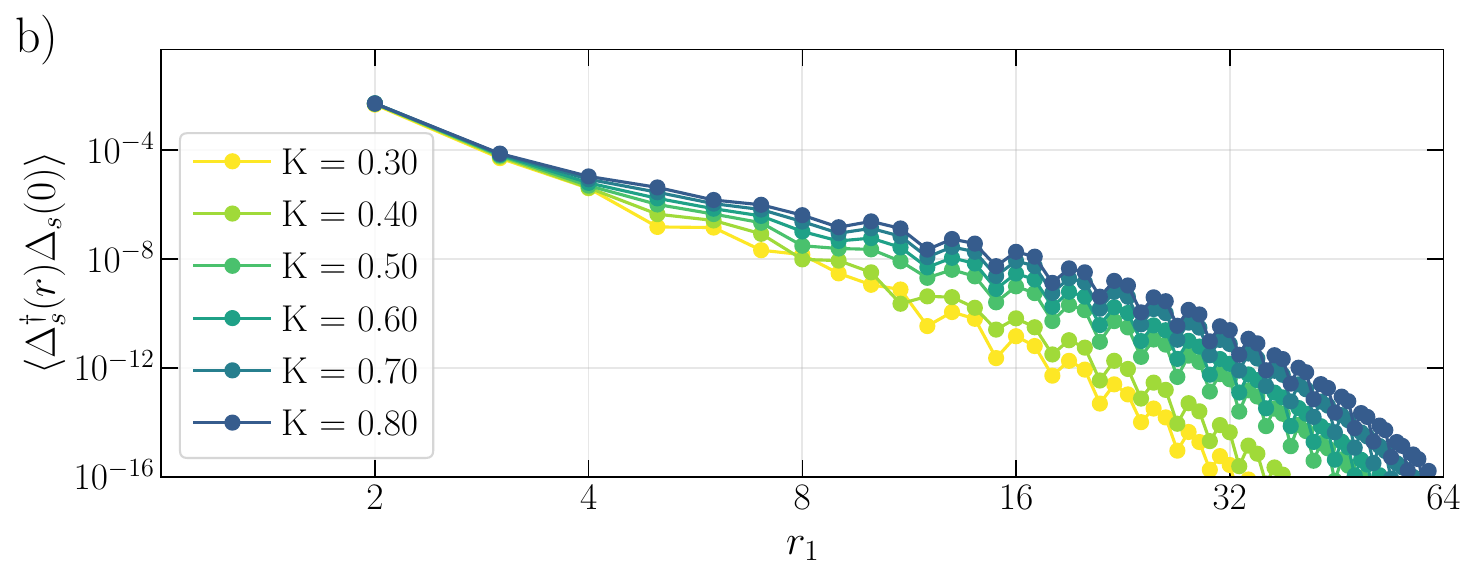}
    \caption{Log-log plot of singlet pair correlator $P_s(\br)$ for doped $K >0 $ Kitaev model 
    with (a)  hole doping $\delta=2/9$ and (b) electron doping $\delta=-2/9$.}
    \label{fig:sc_29}
\end{figure}

\par

In quasi-1D systems, competing SC and charge-density wave correlations with power-law decay, and 
exponentially 
suppressed spin correlations, are characteristic of Luther-Emery (LE) liquids. These are parametrized by
exponents ($K_\rho$, $K_{\rm sc}$), with $K_\rho K_{\rm sc}=1$, which govern the decay
of charge and pair correlators, $\la n({\bf 0}) n({\bf r})\ra \sim r^{-K_{\rho}}$ and
$P_s(\br) \sim \br^{-K_{\rm sc}}$.
We identify the LE liquid as a superconductor when $K_{\rm sc} < 1$. Similar to previous tensor-network studies \cite{Dolfi2015, Jiang2021, Jiang2021b, Peng2021a, Kiely2024}.
We determine $K_{\rm sc}$ by fitting $P_{s}(\br)$ to the form $\br^{-K_{\rm sc}}$ 
over a range of $r$ where it obeys power-law decay, 
before the finite bond-dimension $\chi$ causes an eventual crossover to exponential decay.
We find that the LE state at $\delta  =  2/9$ hole doping, has $K_{\rm sc}  \approx  1.8$, with a negligible dependence on the Kitaev exchange
as seen from Fig. \ref{fig:sc_29}(a).

\begin{figure}
    \centering
    \includegraphics[width=0.995\linewidth]{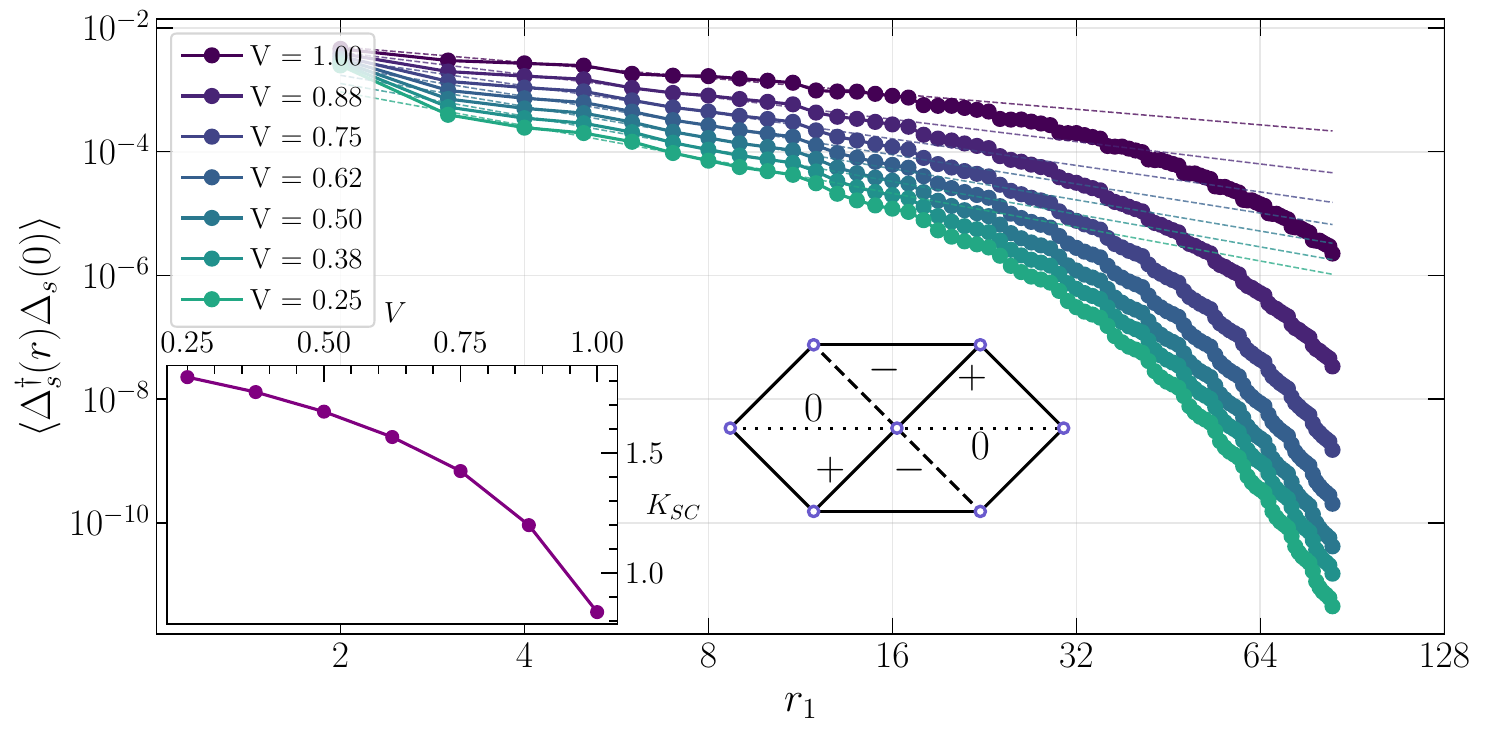}
    \caption{Log-log plot of singlet pair correlations $P_s(\br)$ at hole doping $\delta  =  2/9$ with $K/t =  0.5$ as a function of distance for varying
    nearest-neighbor attraction
    $V$ (with $t=1$). Dashed lines are fits to the form $P_s(\br)  \sim  \br^{-K_{\rm sc}}$.  The inset shows $K_{\rm sc}$ as a function of $V$,
    indicating that SC is stabilized
    for $V/t  \gtrsim  1$ when $K_{\rm sc}  <  1$. The schematic shows the bond pair amplitudes $[-,+,0,-,+,0]$ indicating a
    $d$-wave pairing form factor.}
    \label{fig:v_sc}
\end{figure}

The electron doped state, $\delta = -2/9$, has an even more suppressed $P_s(\br)$ as seen from in Fig. \ref{fig:sc_29}(b), 
but increasing the Kitaev coupling leads to a decrease from $K_{\rm sc}  \approx  3$ to $K_{\rm sc}  \approx  2.5$. Neither state, $\delta = \pm 2/9$ 
is, however, consistent with
superconducting order with $K_{\rm sc}  <  1$. A possible fate of this hole doped state in 2D is a charge density wave or
an algebraic charge liquid, i.e. a spin-gapped holon metal \cite{Kaul2008}.
Going beyond the $t$-$K$ model, we therefore explore whether incorporating a nearest-neighbour attraction term in the Hamiltonian 
$H_V = -V \sum_{\langle \br \br' \rangle} n_\br n_{\br'}$ can favor SC. 
Phonons have been proposed to provide a microscopic origin of such an attractive term 
in various materials \cite{Wang2021b, Qu2022, Wang2025c}.
 As seen from Fig.\ref{fig:v_sc}, $K_{\rm sc}$ decreases with increasing $V/t$, eventually leading to a superconducting state
 with $K_{\rm sc}\approx 0.8$ for $V/t  \approx   1$.
At these substantial interaction strengths, this superconductivity appears primarily due to the attractive interactions.
Nonetheless we can ask whether this SC state inherits any of the anisotropic tendencies from the parent Kitaev magnetic state? 
Our lattice geometry does not support $d+id$ pairing given the reduced point group symmetry. 
We find, instead, that $P_s(\br)$ has a $d$-wave form factor with bond amplitudes $[0,+,-,0,+,-]$ while rotating around a site 
\cite{Koretsune2005}, which is consistent with $d$-wave singlet SC in this geometry.
Up to the $1$D constraints on symmetry, this pairing symmetry is reminiscent of that of the Kitaev SC state emerging from the MFT calculation, although the role of pure attractive interactions \cite{Wang2025c} from one where Kitaev exchange assists this pairing from attraction is not easy to disentangle.
This calls for a more comprehensive study of the SC pairing of a Gutzwiller projected state due to attractive phonon interactions to future work. 
 
\textit{Discussion.---}
We have explored the possibility of SC in the doped Kitaev model on the triangular lattice, a paradigmatic example of a strongly correlated
metal with bond-anisotropic exchange interactions. 
While parton MFT finds large regions of the phase diagram containing topological SC, our tensor network simulations instead uncover emergent spin-charge modulated stripe orders at weak doping which give way to a strongly correlated metal at larger hole doping, with $d$-wave SC emerging in the presence of strong additional attractive interactions.
These results demonstrate the resilience of spin-stripes, such as seen in \cite{Xie2024}, upon doping.
Future work can better explore the fate of these states upon approaching the $2$D thermodynamic limit.
\par
Our study also indicates the importance of kinetic magnetism must be explored in frustrated triangular magnets as well as other lattice geometries \cite{Glittum2025}.
An important future direction would be to extend our study to incorporate all symmetry-allowed exchange terms and further neighbor couplings. 
Such studies may also benefit from exploring the limit of large $t/K$ where the kinetic effects may be explored in greater detail.
Future work may also refine the boundaries of superconductivity upon improving the initial MPS ansatz with other numerical methods \cite{Xu2024} 
In closing, we suggest that hole doping of triangular magnets with antiferromagnetic Kitaev terms, such as NaRuO$_2$ 
\cite{Ortiz2023, Razpopov2023, Bhattacharyya2023} even if it has other exchange terms, would be of immense value to search for 
the predicted spin-charge modulated stripes and unconventional superconductivity. 
This work should also serve as impetus for materials discovery to search for new systems exhibiting dominant bond-anisotropic 
AFM Kitaev interactions on the triangular lattice.
\par

\textit{Acknowledgments.---} We thank A. Bose, S. Divic, E. M. Stoudenmire, L. Devos, and A. Feuerpfeil for helpful discussions.
VUMPS calculations were performed on top of ITensors.jl \cite{ITensor2022}. 
We acknowledge support from the Natural Sciences and Engineering Research Council (NSERC) of Canada via Discovery Grant RGPIN-2021-03214,
and an NSERC Alliance Catalyst Quantum Grant ALLRP 592615-23.
A.H. acknowledges support from a NSERC Canada Graduate Research Scholarship (CGRS-D).
The Flatiron Institute is a division of the Simons Foundation. 
This research was enabled in part by support provided by Compute Ontario (computeontario.ca) and the Digital Research Alliance 
of Canada (alliancecan.ca).
\bibliography{magnets}
\clearpage
\section{Appendices}
\subsection{Parton Mean-Field Theory}

The resulting Hamiltonian after the mean-field decomposition of Eq.\ \ref{eq:gutzwillerhamiltonian} is  
\begin{equation} 
\begin{aligned} H= & \sum_{{\langle \br, \br' \rangle}}- t \left(f_{\br, \s}^{\dagger} f^\pdg_{\br',\s} b^\pdg_\br b_{\br'}^{\dagger} +\text { H.c. }\right) \\ &+\frac{K}{4}\sum_{{\langle \br, \br' \rangle}}\left[\left(f_{\br, \a}^{\dagger} \s^{\a \b}_\mu f^\pdg_{\br', \b}\right)\left(f_{\br,\a}^{\dagger} \s^{\a \b}_\mu f^\pdg_{\br',\b}\right)\right] \\ & +\sum_\br\left[\lambda_\br\left(f_{\br, \s}^{\dagger} f^\pdg_{\br, \s}+b_\br^{\dagger} b^\pdg_\br+1\right)-\mu b^\pdg_\br b_\br^{\dagger}\right] . \label{eq:sb} 
\end{aligned} 
\end{equation}
with a Lagrange multiplier $\lambda_\br$ used to impose the Gutzwiller constraint.
At zero temperature, the bosons condense with a condensate expectation value of 
$
\sqrt{\left\langle b^{\dagger}_\br b^\pdg_{\br + 1}\right\rangle} \approx b^\pdg_\br \approx b_\br^{\dagger} \approx \sqrt{\delta}
$.
 This ansatz remains self consistent when the density order remains uniform. We find this to be true for the $3$ and $4$ site unit cells we consider here. 
In this analysis, we neglect possible phase fluctuations of the condensate along with potential gauge fluctuations from this fractionalization procedure.
This mean-field treatment of the bosons modifies the kinetic term as $t \rightarrow t_f = t\left\langle b_\br^{\dagger} b^\pdg_{\br + 1}\right\rangle = t \delta$. 
This provides the agreement necessary with the Gutzwiller projection that the initial charge and spinon kinetic energy is zero in the insulating regime.  

We retain unbiased Hartree terms for the density and spin orientation at each site as 

$\overline{S}_\mu(\br) =  \langle f_{\br \a}^\dagger \s^{\a \b}_\mu f_{\br \b} \rangle$ with $\m \in {I, x,y,z}$,  $i.e. \left(\overline{n}(\br) = \overline{S}_I(\br)\right)$,  and Fock terms for each bond with $\chi_\mu({\br} ) = \langle f_{\br, \a}^\dagger \s^{\a \b}_\mu f_{\br+1, \b} \rangle $. 
In order to study pairing, we also include decompositions into pairing terms of both singlet and triplet operators, given by $\Delta_s({\br}) = 1/ \sqrt{2}  f_{\br, \a} i \s^{\a, \b}_y f_{\br+1 ,\b} $, 
$
\Delta_{t_1}(\br)=c_{\br, \uparrow} c_{\br+1, \uparrow},  \Delta_{t_2}(\br)=c_{\br, \downarrow} c_{\br+1, \downarrow}
$, and $
\Delta_{t_3}(\br)=\frac{1}{\sqrt{2}}\left(c_{\br, \downarrow}c_{\br+1, \uparrow}+c_{\br, \uparrow} c_{\br+1, \downarrow}\right)
$.
These terms are treated on equal-footing with the regular order parameters. 
The mean-field theory is then numerically solved using root finders until we arrive at self-consistent solutions. 
\subsection{VUMPS Details}  
\begin{figure}[ht!]
\includegraphics[width=0.99\linewidth]{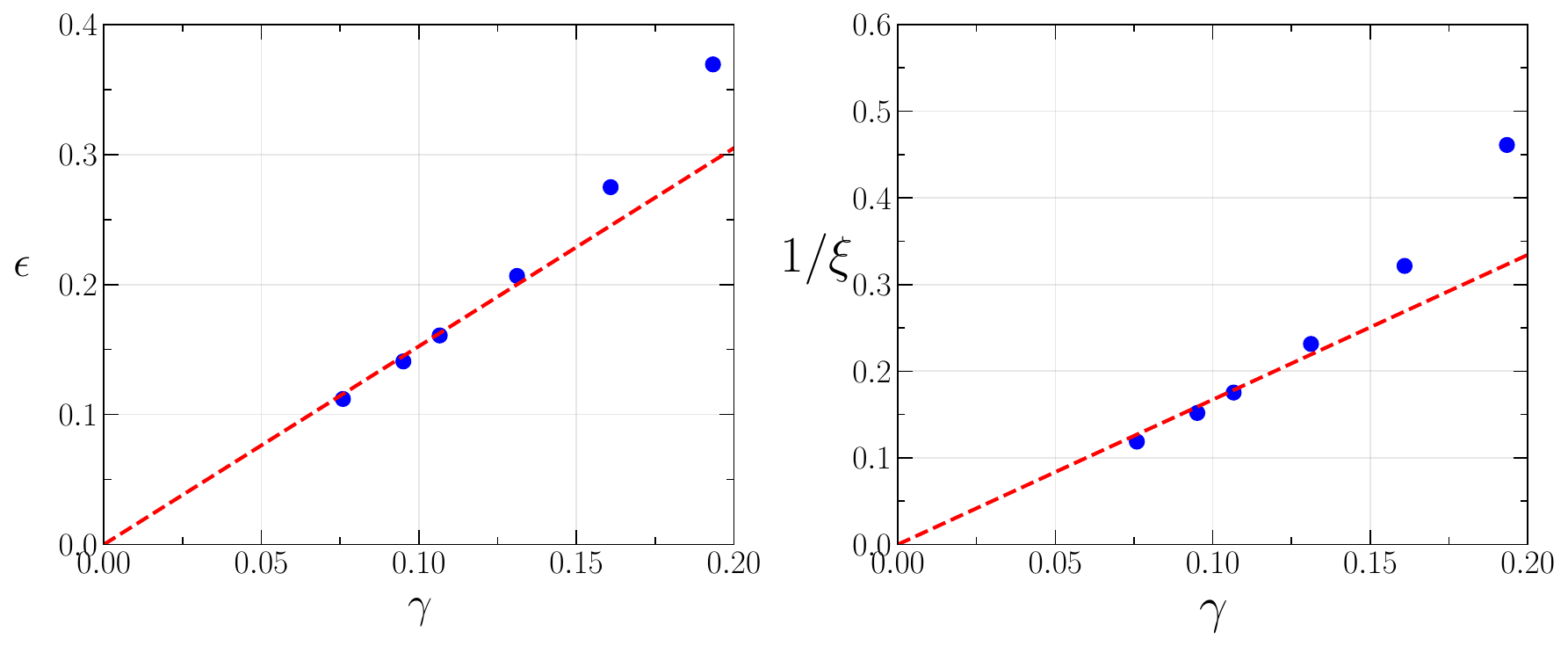}
    \caption{For $K =0.6$ ($t = 1$) and $\delta = 2/9$, the scaling of the transfer matrix eigenvalues show that upon increasing bond-dimension as the system approaches the thermodynamic limit, the correlation length $\xi$ diverges and the transfer matrix gap $\e = 1 - \epsilon_1$ begins to admit a linear scaling towards $\e \rightarrow 0$, a gapless state. }
    \label{fig:scaling}
\end{figure}

Due to the entanglement scaling of matrix product states (MPS) \cite{Cirac2021}, the VUMPS algorithm is limited to quasi-one dimensional systems with a finite number of `legs', or `width' in the transverse direction.
Within this framework, the space of UMPS for a given bond dimension, $\chi$, is interpreted as a manifold with a corresponding tangent space structure \cite{Haegman2014}.
By considering gradient descent computed in this tangent space, one can devise a variational optimization algorithm to find the UMPS with the lowest energy for a given physical system \cite{Zauner-Stauber2018}.  
As such, the VUMPS algorithm works at fixed bond dimension. 
In order to increase the bond dimension, we use a controlled bond dimension expansion procedure with the time-dependent variational principle (TDVP) \cite{Haegeman2011, Gleis2022, Gleis2023, Li2024}. \par
We caution however that as is the case with global variational searches, VUMPS can be sensitive to initial wave-function guesses or become stuck in local minima.
All calculations are done on an triangular `cylinder', with periodic (infinite) boundary conditions in the $L_2$ ($L_1$) direction up to a maximum $\chi = 2400$.
The convergence is done with a Schmidt tolerance $\e_{\mathrm{SVD}} = 10^{-5}$.
This corresponds to the energy converging to $\Delta E \leq  10^{-9}$. 
Tensor network methods when dealing with gapless states effectively gap the state due to a finite bond-dimension, providing signatures of a diverging correlation length.
We utilize these insights to perform a scaling fit of the dominate eigenvalue of the transfer matrix $  \xi = - 1/\log(\e_1) $ as a function of the refinement parameter $\g = \e_1 - \e_2$ in Fig. \ref{fig:scaling}. Through the procedure detailed in \cite{Rams2018}, we are able to detail the gapless nature of the $\delta = 2/9$ doped state as the correlation length diverges.  

\subsection{Charge Correlation Data}
We provide additional plots here for charge correlations upon doping for $\delta = 1/9$ and $2/9$ of hole and electron doping.
\onecolumngrid
\begin{figure*}
     \centering
        \includegraphics[width=0.49\textwidth]{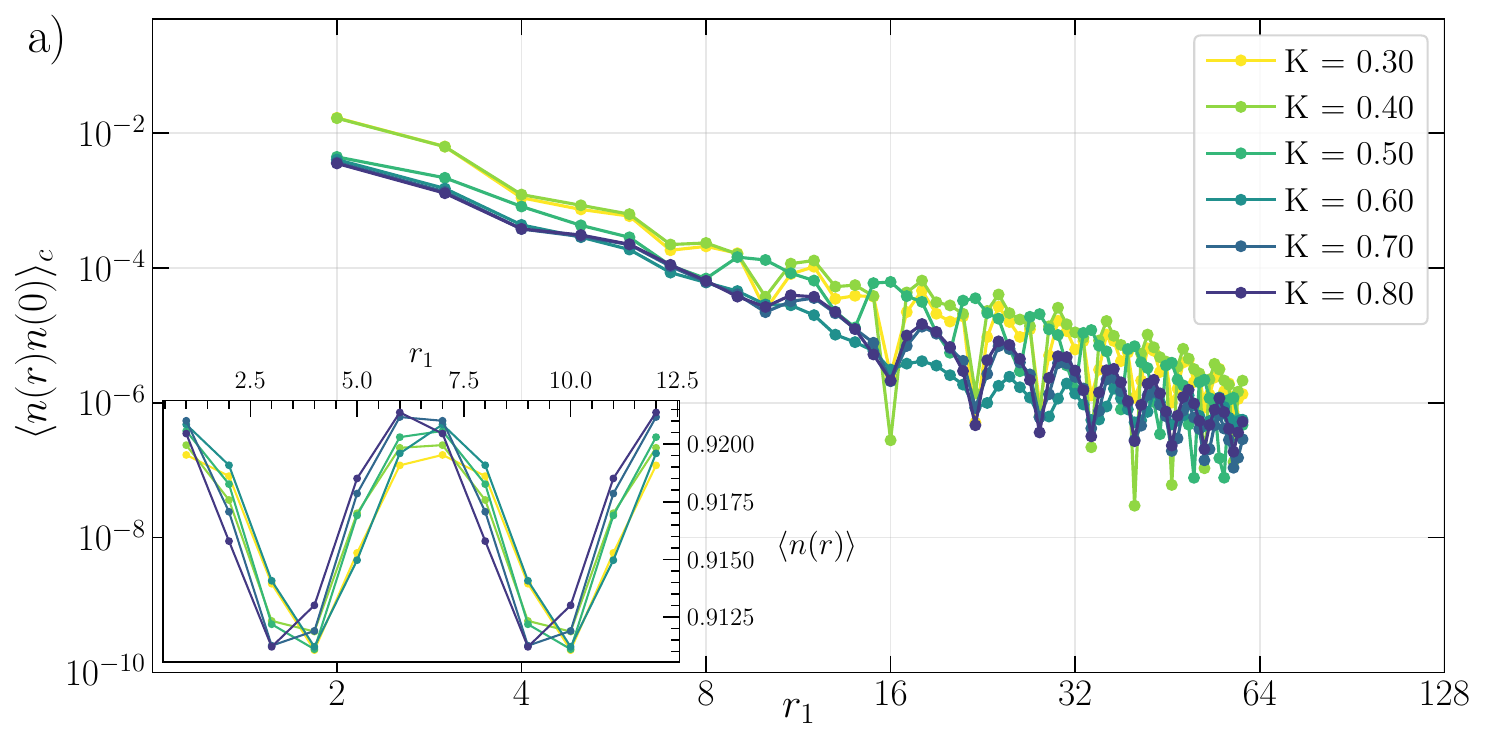}
        \includegraphics[width=0.49\textwidth]{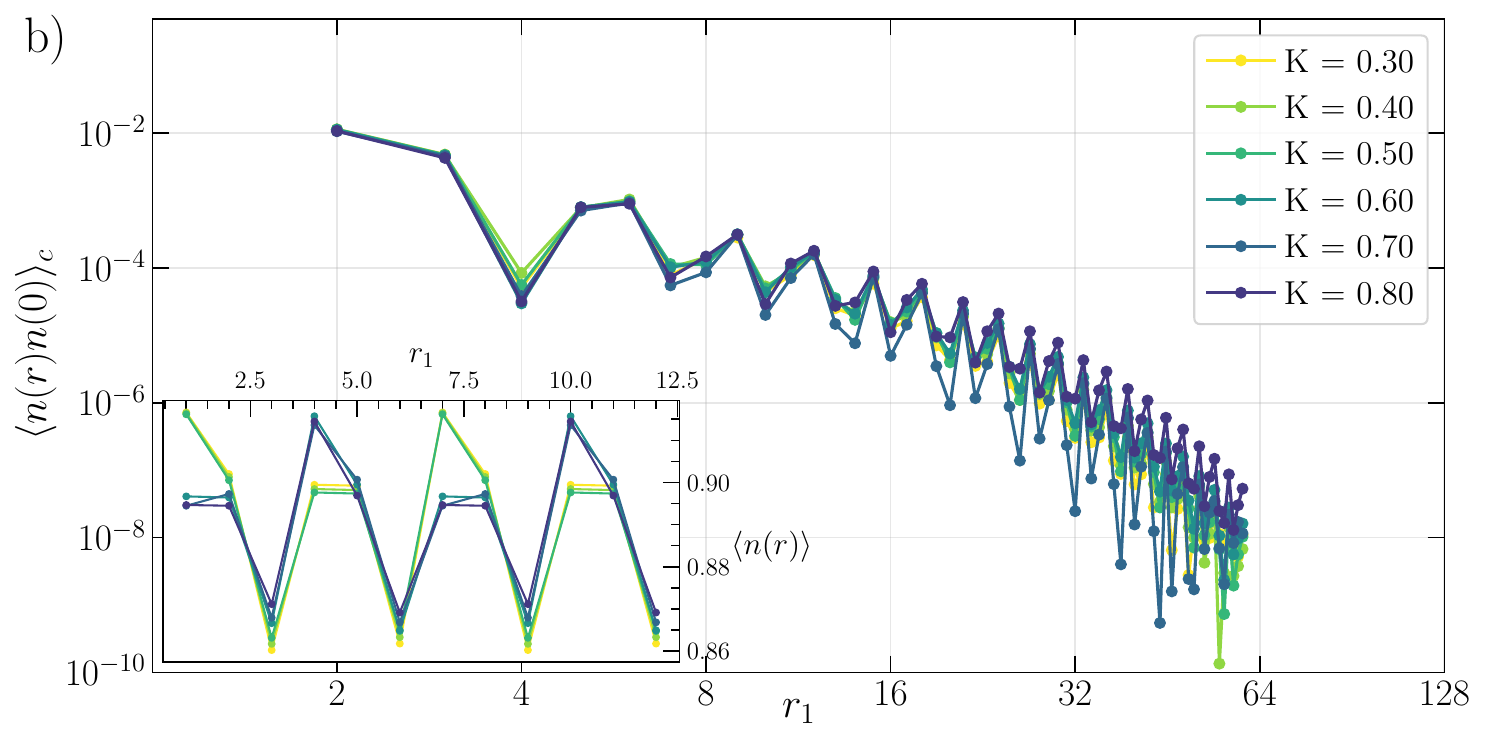}
    \caption{ $N(\br)$ connected correlation function with local $\langle n(\br) \rangle$ expectation as inset for the $L_2 = 3$ state at  $\delta = 1/9$ a)   hole doping and b) electron doping  of the $K >0$ ($t = 1$) Kitaev model.}
    \label{fig:charge_19}
\end{figure*}
\begin{figure*}
     \centering
\includegraphics[width=0.49\textwidth]{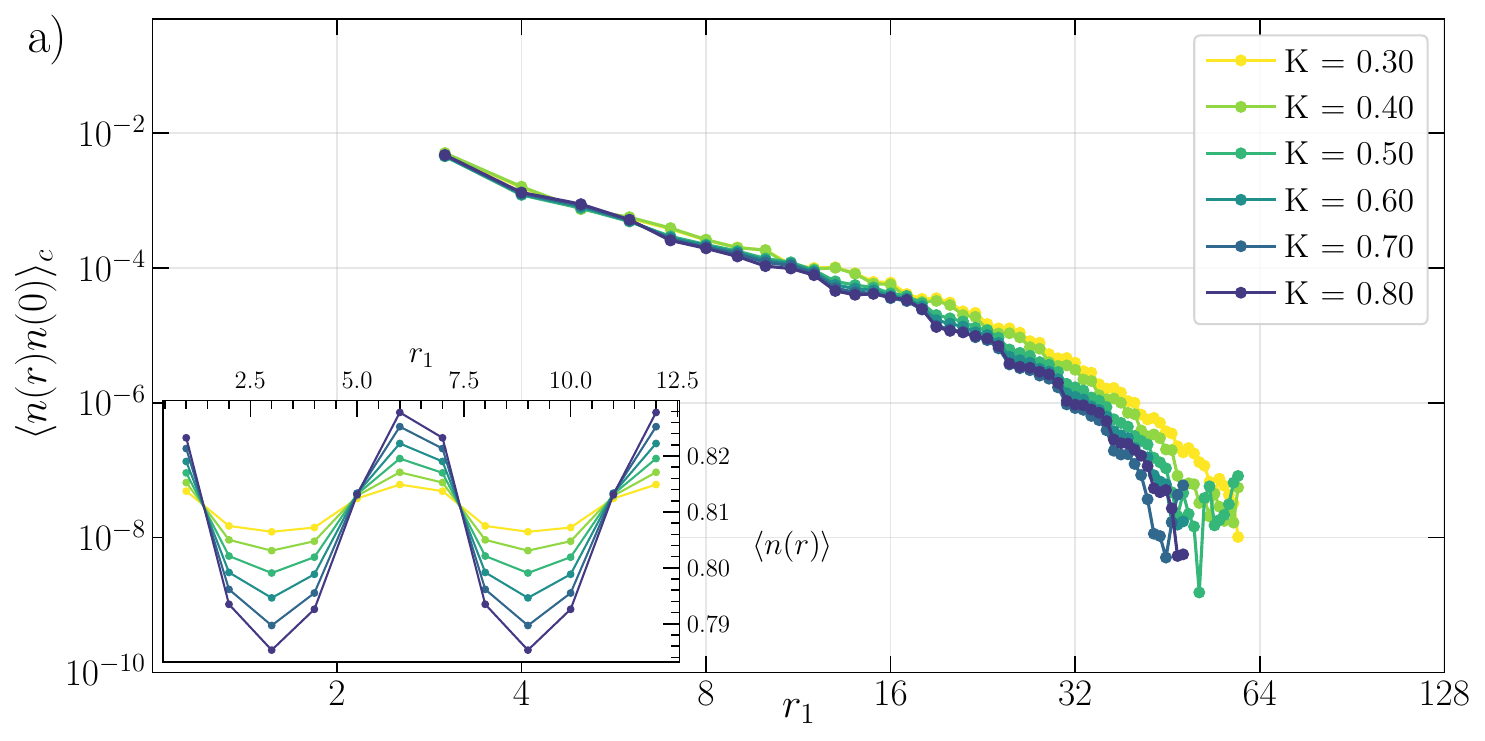}
\includegraphics[width=0.49\textwidth]{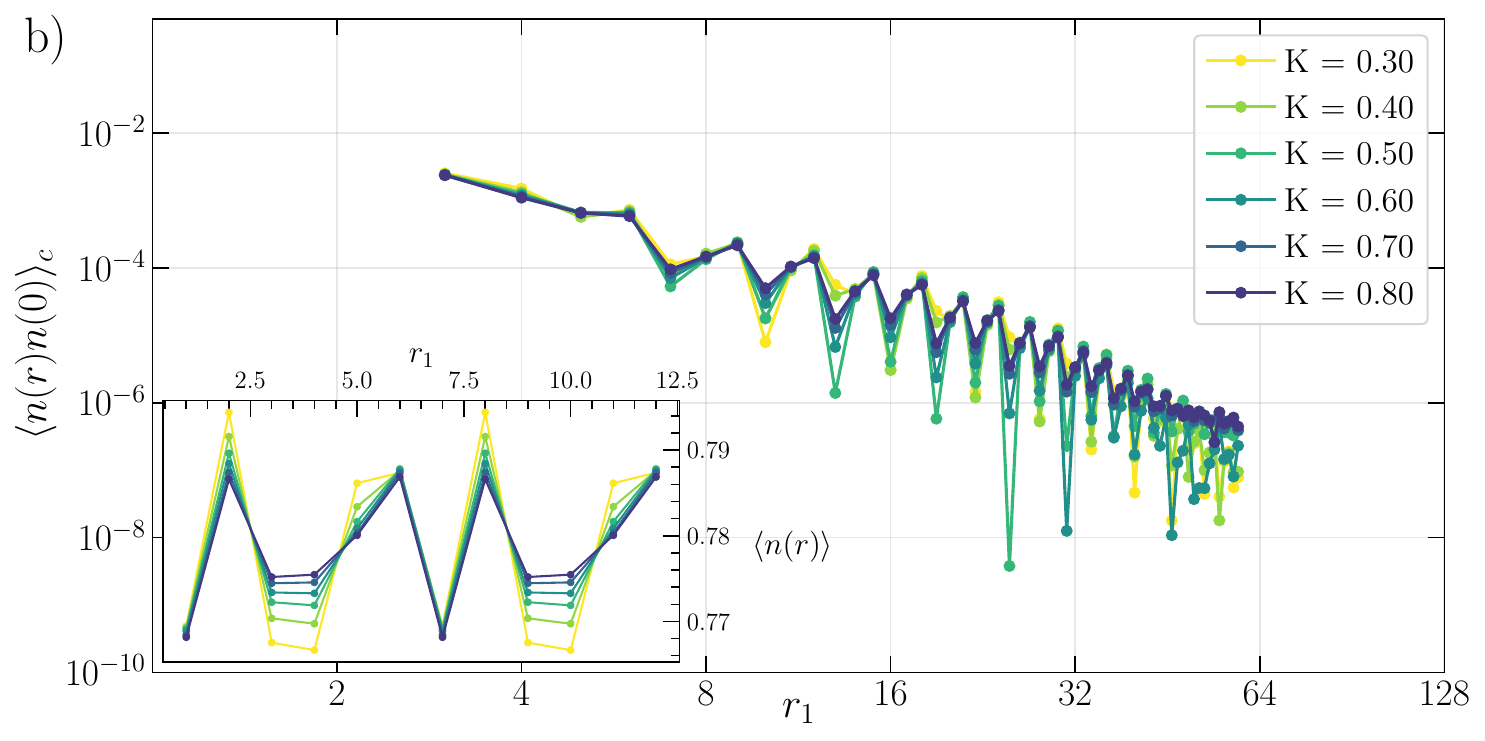}
    \caption{ $N(\br)$ connected correlation function with local $\langle n(\br) \rangle$ expectation as inset for the $L_2 = 3$ state at  $\delta = 2/9$ a)   hole doping and b) electron doping  of the $K >0$ ($t = 1$) Kitaev model.}
    \label{fig:charge_29}
\end{figure*}
\setlength{\belowcaptionskip}{-10pt}
\clearpage

\newpage
 

\section*{Supplementary material}

# Supplementary Material for:
# Hunting for superconductivity in doped triangular lattice Kitaev magnets

Andrew Hardy,[1] Ryan Levy,[2] and Arun Paramekanti[1]

[1]*Department of Physics, University of Toronto,
60 St. George Street, Toronto, ON, M5S 1A7 Canada*

[2]*Center for Computational Quantum Physics, Flatiron Institute, New York, New York 10010, USA*

(Dated: August 21, 2025)

## I. TRIPLET AND FERROMAGNETIC CORRELATIONS

We provide additional details about triplet and ferromagnetic correlations. we find that for $K < 0$, as mentioned in the main text, the state exhibits suppressed triplet pairing and resilient $\mathbf{Q}=0$ ferromagnetism. The $\delta = \pm 1/9$ states displayed in Fig. 1 shows strong ferromagnetic order for both hole and electron doping. The net magnetization is also a larger value than that on the $(K > 0)$ AFM side, indicative of the less frustrated nature of the ferromagnetic exchanges.

This order persists to the $\delta = \pm 2/9$ states, albeit with some modulation of the magnetization, indicated in Fig. 2, The magnetic order suppresses the SC, as shown in the exponential decay of $P_t(\mathbf{r})$, the triplet correlation functions, show in Fig. 3. We remark in passing that $P_t(\mathbf{r})$ is also sharply reduced for the $\delta = \pm 2/9$ AFM states, show in Fig. 4, as expected. These results in turn motivates our focus on $K > 0$ and singlet correlations.

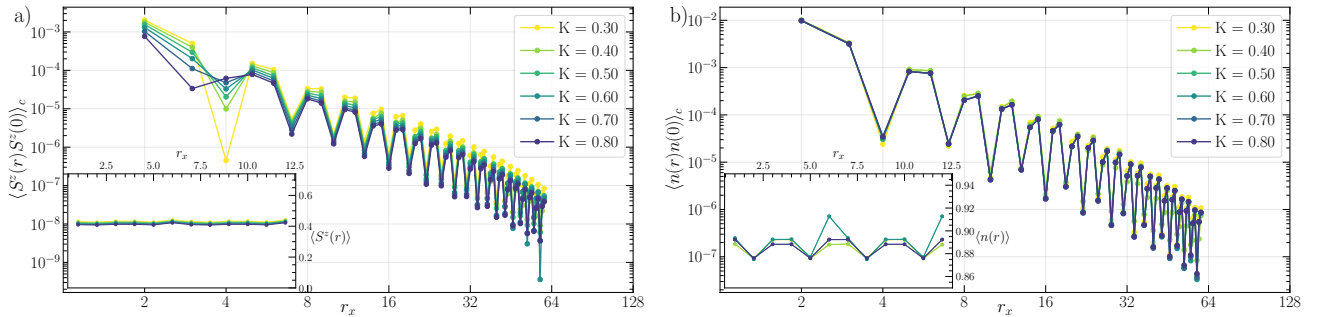

FIG. 1: a) $S(\mathbf{r})$ connected correlation function with local $\langle S^z \rangle$ expectation as inset and b) $N(\mathbf{r})$ connected correlation function with local $\langle n(\mathbf{r}) \rangle$ expectation as inset for the $L_1 = 3$ state at $\delta = 1/9$ electron doping of the ferromagnetic $K < 0$ $(t=1)$ Kitaev model.

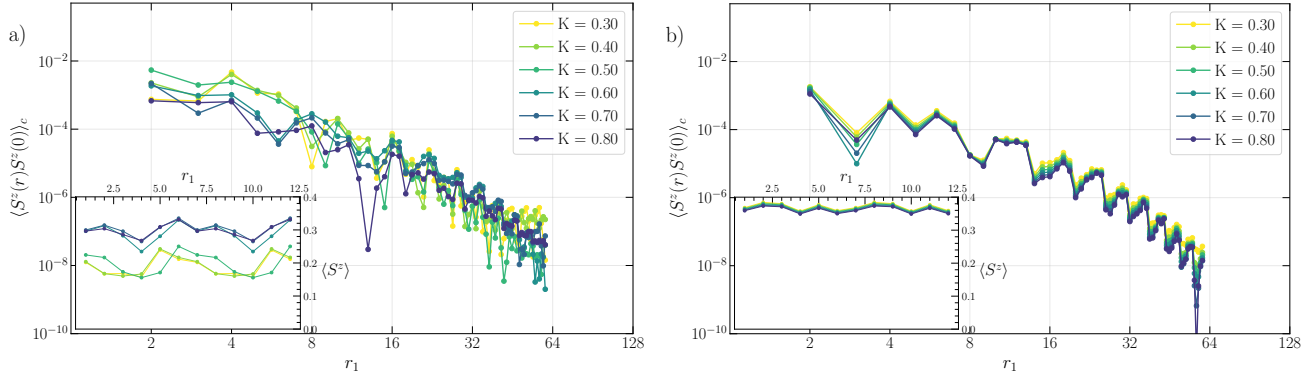

FIG. 2: $S(\mathbf{r})$ connected correlation function with local $\langle S^z \rangle$ expectation on the first leg as inset for the $L_1 = 3$ state at $\delta = 2/9$ a) hole doping and b) electron doping of the $K < 0$ ($t = 1$) Kitaev model.

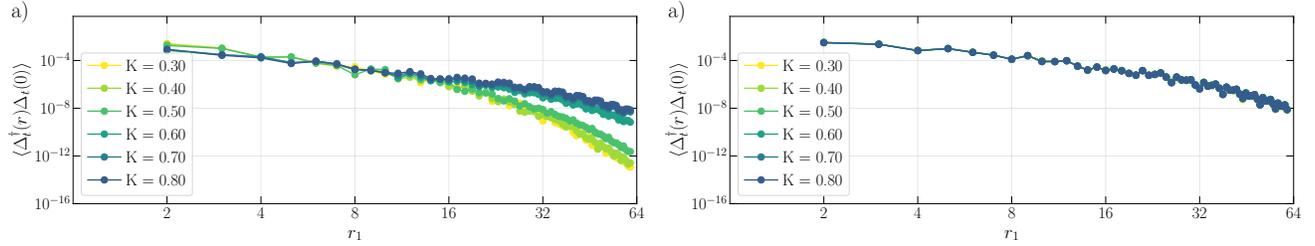

FIG. 3: $P_t(\mathbf{r})$ superconducting triplet correlations for the $L_1 = 3$ state for $\delta = 2/9$ doping of the $K < 0$ ($t = 1$) Kitaev model with a) hole doping and b) electron doping. The local expectation values of these are zero in a $U(1)$ conserved state.

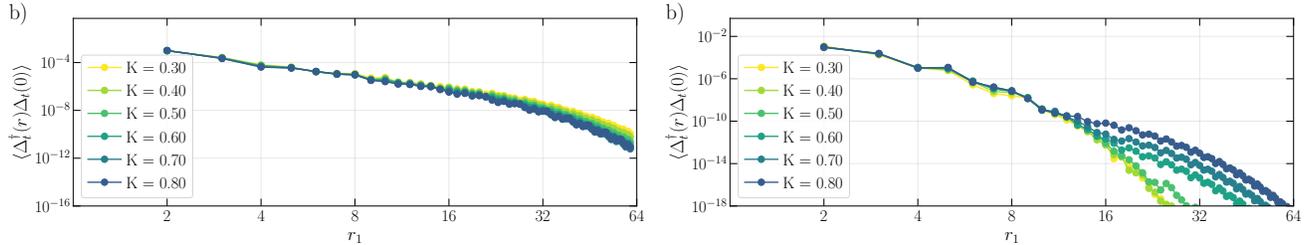

FIG. 4: $P_t(\mathbf{r})$ superconducting triplet correlations for the $L_1 = 3$ state for $\delta = 2/9$ doping of the $K > 0$ ($t = 1$) Kitaev model with a) hole doping and b) electron doping. The local expectation values of these are zero in a $U(1)$ conserved state.

 
\end{document}